# A Short-Term Integrated Wind Speed Prediction System Based on Fuzzy Set Feature Extraction

Yijun Geng, Jianzhou Wang*, Jinze Li, Zhiwu Li

***Yijun Geng***: School of Statistics, Dongbei University of Finance and Economics, Dalian, 116025, China
E-mail address: 2022211008@stumail.dufe.edu.cn

*** Corresponding author: Jianzhou Wang***: School of Statistics, Dongbei University of Finance and Economics, Dalian, 116025, China
E-mail address: wjz@lzu.edu.cn.

***Jinze Li***: School of Statistics, Dongbei University of Finance and Economics, Dalian, 116025, China
E-mail address: 13147872112@163.com

***Zhiwu Li***: Institute of Systems Engineering, Macau University of Science and Technology, Macao, 999078, China
E-mail address: zwli@must.edu.mo.

**Abstract**

Wind energy has significant potential owing to the continuous growth of wind power and advancements in technology. However, the evolution of wind speed is influenced by the complex interaction of multiple factors, making it highly variable. The nonlinear and nonstationary nature of wind speed evolution can have a considerable impact on the overall power system. To address this challenge, we propose an integrated multiframe wind speed prediction system based on fuzzy feature extraction. This system employs a convex subset partitioning approach using a triangular affiliation function for fuzzy feature extraction. By applying soft clustering to the subsets, constructing an affiliation matrix, and identifying clustering centers, the system introduces the concepts of inner and boundary domains. It subsequently calculates the distances from data points to the clustering centers by measuring both interclass and intraclass distances. This method updates the cluster centers using the membership matrix, generating optimal feature values. Building on this foundation, we use multiple machine learning methods to input the fuzzy features into the prediction model and integrate learning techniques to predict feature values. Because different datasets require different modeling approaches, the integrated weight-updating module was used to dynamically adjust model weights by setting a dual objective function to ensure the accuracy and stability of the prediction. The effectiveness of the proposed model in terms of prediction performance and generalization ability is demonstrated through an empirical analysis of data from the Penglai wind farm.

## 1. Introduction

Recently, many countries have been grappling with energy shortage, which impedes global sustainable development. The sharp rise in global energy prices has posed a serious threat to industries in both Asian and European nations, with the potential for power outages becoming a significant concern. As the global energy crisis deepens, renewable energy is increasingly recognized as a crucial solution. As a prominent example of renewable energy, wind energy is receiving increasing attention and investment from many countries and regions owing to its advantages of abundant resources, environmental friendliness, and zero emissions (Wang et al., 2023).

Accurate wind speed forecasting is essential for optimizing wind energy generation. This forecasting plays a key role in ensuring grid stability and continuous power supply while also minimizing operational costs and improving turbine control strategies. The accuracy of wind speed forecasting directly impacts turbine blade pitch adjustments and the optimization of maximum wind power tracking. However, the inherent randomness, intermittency, and variability of wind resources make wind speed forecasting a complex and challenging task, often resulting in wind curtailment and power losses (Bentsen et al., 2023).

To improve the accuracy of wind speed forecasting, researchers have developed various innovative methods, which can be categorized into physical, statistical, and artificial intelligence approaches. Physical forecasting models rely on meteorological data from numerical prediction systems, along with factors such as geographic features, surrounding conditions, and site-specific configurations. These models simulate atmospheric states, fluid dynamics, and thermodynamic processes by computing complex mathematical equations to comprehensively analyze changes in meteorological elements, thus predicting wind speed. However, this approach has notable limitations, primarily because the accuracy of these models heavily depends on the precision of the initial conditions. Because accurately measuring and capturing the initial conditions of an atmospheric system are challenging, errors tend to accumulate and propagate, consequently reducing the accuracy of the forecasts.

To address the limitations of traditional physical methods in wind speed prediction, researchers have increasingly turned to statistical models to enhance prediction accuracy. Common data analysis–based methods include the persistence method and time series forecasting models. The persistence method is simple and requires minimal data computation, generally using the current wind speed value to predict the wind speed at the next moment, making it suitable for short-term forecasts. By contrast, time series forecasting models rely on historical data, constructing more refined wind speed forecasting models by analyzing the time series characteristics and autocorrelations of wind speed. These methods can capture the regular patterns and trends in wind speed variations, making them suitable for wind speed forecasts over a long period. For example, Li et al. (2021) proposed an innovative Markov prediction method for integrating wind speed data. This method converts wind speed data into a discrete state sequence using a QR codebook, which is linked to the joint distribution of speed and acceleration. The discrete state sequence is then used to construct a state transition probability matrix, resulting in improved prediction accuracy compared with that provided by traditional statistical models. Aasim et al. (2019) introduced a Db2-based method for applying the maximum overlap discrete wavelet transform to wind speed, effectively identifying the timing of high-frequency components and efficiently analyzing low-frequency variations. Moreover, their novel repeated WT-based ARIMA (RWT-ARIMA) model decomposes high-frequency wind speed variations into finer levels. The autoregressive integrated moving average model (ARIMA) model was then employed to model these decomposed high-frequency wind speed variations. However, as a statistical approach, time series forecasting requires a significant amount of historical wind speed data, which is often difficult to obtain accurately from many wind farms.

In today's era of big data, alongside the rapid expansion of data volumes, traditional statistical and mathematical models face significant challenges. These models often struggle to manage large-scale data analysis and identify key factors in predictive datasets. To address these challenges, researchers have turned to artificial intelligence models. Specifically, optimizing neural network algorithms to improve prediction accuracy has become a prominent area of research. Additionally, techniques for data denoising and enhancement algorithms have been widely explored and applied within the academic community. Moreover, some statistical models are only effective in specific regions or under certain conditions, making them difficult to generalize to other areas or different meteorological environments.

Recently, researchers have proposed several hybrid methods for wind speed prediction. These methods typically combine various models and algorithms to enhance the accuracy and reliability of forecasts. For instance, Yu et al. (2023) introduced a model using an attention mechanism, with the graph attention network (GAT) serving as the main framework for spatial feature extraction. This model effectively aggregates and extracts wind speed data from both the target station and nearby stations. The spatiotemporal features extracted by the GAT are then fed into the Informer, which trains the prediction framework to generate wind speed forecasts for the target station. Sarangi et al. (2023) proposed a method using variational mode decomposition (VMD) to decompose primary time series data into several modes. A combination of Gauss–Bernoulli restricted Boltzmann machine (GBRBM) and Bernoulli–Bernoulli RBM was then used within the deep belief network. In this framework, the GBRBM serves as the initial RBM, converting the continuous attributes of the source data into binary distribution features, thus making this approach suitable for probabilistic wind speed forecasting. Xu et al. (2023) developed an innovative wind speed forecasting model using phase space reconstruction and a broad learning system (BLS). Phase spaces were reconstructed using different delay dimensions and phase scales, and a natural neighborhood spectrum was created from the phase vector, facilitating the selection of the optimal phase space without parameter tuning. This step determined the optimal input quantity for the BLS. Elastic net regularization was incorporated to reduce overfitting, and BLS was incrementally trained. Pelaez-Rodriguez et al. (2023) introduced a hierarchical framework that partitions training data into fuzzy-soft clusters based on target variable values. This integrates predictions from individual models into a fuzzy-based system and uses a differential evolution optimization algorithm to determine the optimal data partitioning method. For the final regression scheme, a fast-training stochastic neural network approach is employed. Bommidi and Teeparthi (2024) proposed another hybrid forecasting method that combines an adaptive denoising method with an autoformer model, demonstrating the model's sophistication by evaluating it with wind speed data from wind farms. Sareen et al. (2023) presented a combinatorial model using the denoising autoencoder algorithm, incorporating the VMD algorithm with the bidirectional long short-term memory deep learning algorithm to enhance prediction accuracy. Wang et al. (2023) introduced a multimodal model based on integrated learning for ex officio wind speed prediction by reconstructing the time series, splicing the subsequences, and integrating them. Parri et al. (2023) proposed the VMD-Ts2Vec-SVR method, which uses the contextual time series representation (Ts2Vec) model and support vector regression (SVR) for data extraction and wind speed prediction after denoising using VMD. The outstanding performance of this method across all time scales is demonstrated through two experiments.

**Table 1** presents the results of existing research. While the aforementioned literature demonstrates substantial improvements in wind speed forecasting compared with statistical and physical methods, several issues remain in current research:
1. Most data preprocessing methods currently rely on VMD techniques to separate data into different frequency components, typically treating the highest frequencies as white noise to be eliminated. However, this approach lacks a solid theoretical foundation, which can lead to incomplete noise removal or the unintentional elimination of relevant non-noise components.
2. Relying on a single forecasting method can compromise the stability and accuracy of predictions. This is particularly true when appropriate optimization measures are lacking, which considerably limits the effectiveness and applicability of a single approach.
3. An overly complex model may result in overfitting, where the model performs exceptionally well on the training dataset but fails to generalize effectively to the test dataset.

To address the limitations of historical wind speed prediction models, this study proposes a novel hybrid wind speed prediction framework, CGT-BF, aimed at overcoming the challenges in wind speed prediction. To the best of our knowledge, this study is the first to introduce a feature extraction method (FIC-MG), which employs a convex subset partitioning technique based on a triangular affiliation function for fuzzy feature extraction. The module is used to soft cluster the data subsets to construct an affiliation matrix and clustering centers while introducing the concepts of inner and boundary domains. These concepts, combined with measurements of interclass and intraclass distances, are used to compute the distances between data points and clustering centers. In the feature prediction module, we propose an innovative integrated learning system known as the hybrid prediction system (T-BF). This system leverages neural networks based on different principles for data prediction, particularly through the proposed LSTM–XGB model. To prevent overfitting, L1 and L2 regularization techniques are applied to constrain the weights. L1 regularization introduces sparsity by eliminating irrelevant features, causing some weights to converge to zero, while L2 regularization reduces anomalous fluctuations by smoothing the weights, thereby ensuring the robustness of the model. Additionally, this study introduces an improved multiobjective sunflower optimization (IMOSFO) algorithm to dynamically weight the integrated model. The optimization algorithm combines tent mapping with a perturbation strategy, generating complex pseudorandom sequences through a nonlinear dynamical system. Despite its simple structure, this optimization method captures chaotic properties such as initial value sensitivity and ergodicity, providing tent mapping with a substantial advantage in global search and effectively avoiding local optimality, thereby enhancing the global exploration ability of the algorithm. The overall structure of the experiment is shown in **Fig. 1**. The key innovations presented in this study are as follows:

1. We propose the FIC-MG method, an innovative data preprocessing approach that combines fuzzy information granulation and fuzzy rough C-means clustering. By applying soft clustering to the subsets, the method constructs an affiliation matrix and clustering centers. It introduces the concepts of inner and boundary domains, which when combined with interclass and intraclass distance measurements can compute the distances between data points and clustering centers. The method updates the cluster centers based on the membership matrix and generates optimal feature values, effectively extracting fuzzy set features and improving data feature extraction efficiency and accuracy.

2. We introduce a dual-framework system to address the issue of ignoring wind speed uncertainty single-use point wind speed predictions. This system comprises a point prediction framework and an interval prediction framework, which complement each other. Traditional wind speed prediction methods primarily focus on point prediction; however, the uncertainty of point prediction is often insufficient when dealing with complex meteorological conditions. To compensate for this deficiency, we introduce an interval prediction framework, which quantifies the uncertainty of wind speed by constructing confidence intervals. These intervals define upper and lower bounds that encompass the true wind speed, providing both a predicted value and an expected range of wind speed fluctuations. The framework can adjust the width of the prediction intervals to reflect varying degrees of uncertainty.

3. We develop a multivariate intelligent granular combination prediction system (CGT-BF) to address the challenge of chaotic time series prediction. This system uses information granulation to accurately predict chaotic time series. It expands the single variable of the time series into a multidimensional space and transforms it into a granular interval. This reconstruction effectively captures the nonlinear dynamic properties of the time series. The data are then decomposed into multiple information granules, each corresponding to a specific time window, and combined prediction results are generated based on the extracted granular data.

4. We propose the enhanced IMOSFO algorithm, which is proven to be effective in obtaining Pareto-optimal solutions from a theoretical perspective. The IMOSFO algorithm optimizes point prediction and granularity in two phases. In the first stage, IMOSFO conducts global search optimization via a mapping mechanism that initializes the spatial distribution of solutions. In the second stage, IMOSFO introduces an adaptive distribution strategy that dynamically adjusts the search step size and direction based on the current distribution characteristics of the solutions, thereby adapting to the prediction needs of different scenarios.

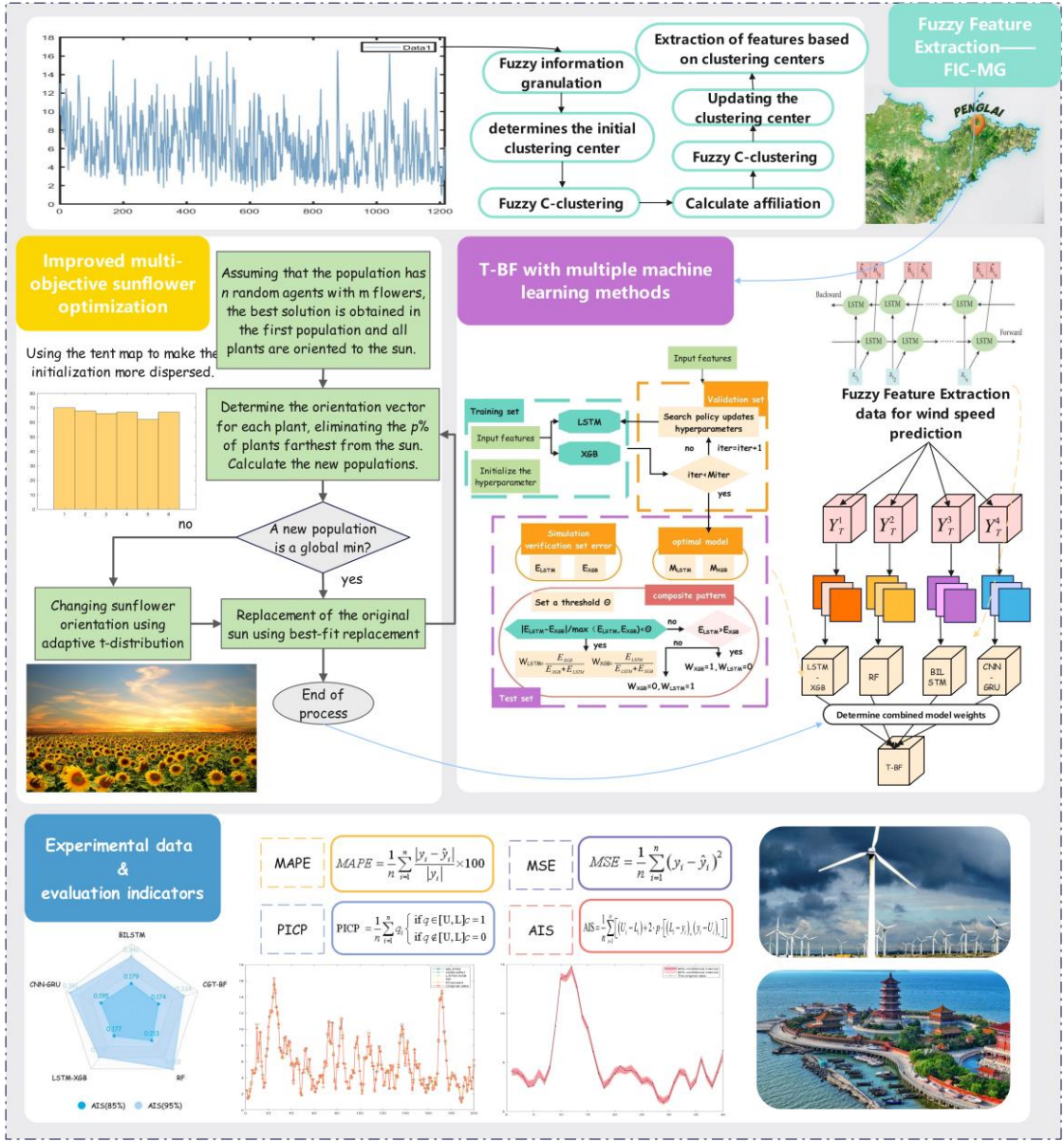

**Fig. 1 Flowchart of the entire paper**

## 2. Experimental methods and data

In this section, we provide a detailed description of the proposed prediction system CGT-BF. The system is composed of the FIC-MG fuzzy feature extraction method and the innovative ensemble learning method T-BF. T-BF integrates several core machine learning methods, including BiLSTM, CNN–GRU, LSTM–XGB, and RF, and utilizes an enhanced optimization algorithm, IMOSFO, to dynamically update the weights.

Table 1

**Literature on wind speed prediction modeling**

| Models | Author | Factors | Conclusions | Advantages | Disadvantages |
|---|---|---|---|---|---|
| Traditional statistical methods | | | | | |
| WS | Li et al. | Wind speed | Incorporating Markov chains into WS to improve prediction accuracy. | Models are more interpretable and are often simpler and easier to implement. | Poor prediction results for chaotic time series. |
| ARIMA | Aasim et al. | Wind speed | ARIMA was used to depict long-term trends in wind speeds. | | |
| Artificial intelligence model | | | | | |
| DBN-MKRVFLN | Sarangi et al. | Wind speed | DBN-MKRVFLN, which uses narrower prediction intervals to bring higher coverage probabilities. | Combining the advantages of multiple methods, the resulting combined model has a high degree of accuracy. | The use of optimization algorithms to optimize the model may lead to long experimentation times and require a large amount of data to support it. |
| GAT-Informer | Yu et al. | Wind speed | GAT-Informer allows for more stable prediction results. | | |
| BLS | Xu et al. | Wind speed | Higher prediction accuracy is achieved by BLS for wind speed prediction. | | |
| EWS | Rodriguez et al. | Wind speed | EWS has a better performance in dealing with extreme wind speed predictions. | | |
| VMD-Ts2Vec-SVR | Parri et al. | Wind speed | VMD-Ts2Vec-SVR achieves the best performance indicators for both wind farms in all time frames. | | |

## 2.1 Data description and processing

A set of 21 wind speed measurements from Penglai, Shandong Province, China, were selected in this study. Three groups of wind speed measurements were randomly chosen as experimental data, named Data1, Data2, and Data3. The experimental data were recorded at 10-min intervals. For the missing data, linear interpolation was applied for imputation. Linear interpolation is a method used to estimate the missing values of one-dimensional data using the values of neighboring data points on either side of the missing point. Although the result of linear interpolation at the central point is equivalent to the mean, it effectively assigns weights depending on the distance to the neighboring data points. For wind speed prediction, a window size of 36 was used with the FIC-MG feature extraction method to obtain wind speed feature data for each wind speed point. Fuzzy information granulation was applied to the wind speed data with the same window size of 36 to obtain same-frequency features and wind speed data. The processed raw data are shown in **Fig. 2**.

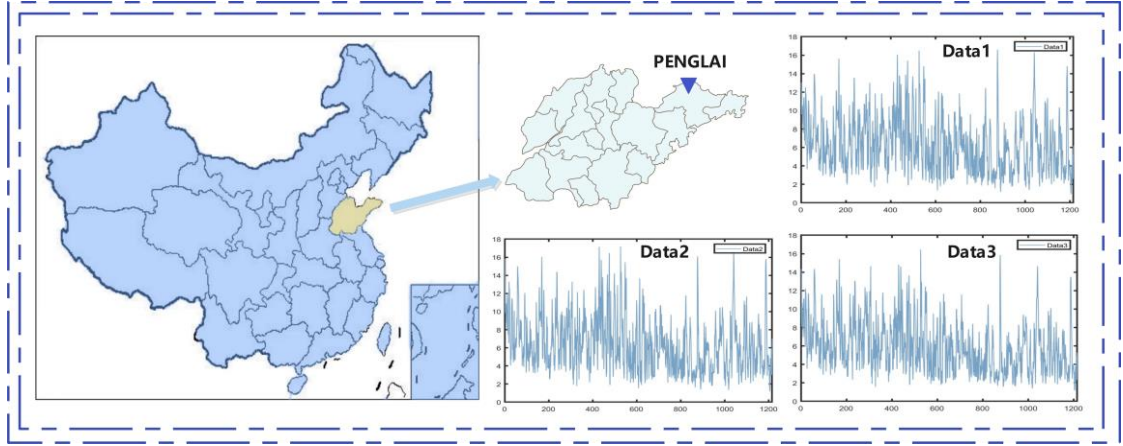

Fig. 2 Experimental data on granulation

## 2.2 Data preprocessing methodology (FIC-MG)

Another data preprocessing method, FIC-MG, has been proposed, which combines fuzzy information granulation and fuzzy rough sets. The core concept of this method is to process the data through information granulation, generating granular intervals that facilitate the extraction of data features to accurately predict wind speed trends and variations. The method involves three stages: fuzzy information granulation, fuzzy rough $C$-clustering, and cluster center updating.

Fuzzy sets, introduced by Professor Zadeh, utilize a "membership function" to express fuzziness. This development marked a breakthrough in Cantor's classical set theory, enabling mathematical approaches to handle fuzzy phenomena. This breakthrough laid a solid foundation for the development of fuzzy engineering. The basic idea of fuzzy sets is to expand the membership relationship of classical sets, broadening the membership degree of elements from the binary values of 0 or 1 to a range of [0,1]. This extension allows for the quantitative characterization of fuzzy concepts and enables objective regularity analysis and research of fuzzy objects (Zheng et al., 2023).

### A. Fuzzy information granularity

The concept of fuzzy information granulation involves dividing an entire dataset into multiple parts, where each part is a subset that exhibits similar characteristics. Fuzzy information granulation extends the prediction time span, which is particularly valuable for analyzing patterns and trends over a broad time range. Fuzzy information granularity is defined as follows:

$$C \triangleq (t\ is\ G), t \in T \qquad (1)$$

where $T=(t_1,t_2,\cdots,t_n)$ represents the original time series and a fuzzy information granule, $C$, is constructed to accurately describe the fuzzy concept $G$ (a fuzzy set with $T$ as the universe of discourse). Once $G$ is determined, the fuzzy information granule is accordingly defined.

Window partitioning involves dividing the time series into multiple subsequences based on specific rules and determining the optimal window size. Herein, a window size of 36 is used, extending the original 10-min interval time series into a 6-h time span.

Information fuzzification refers to the process of applying fuzzy sets to all windowed data to create appropriate fuzzy information granules. Fuzzification enables the granulation results to more effectively capture the relevant information from the data. The goal of this process is to establish a fuzzy particle $C$ on $T=(t_1,t_2,\cdots,t_n)$, which comprehensively describes the fuzzy concept of $T$, and thus determines the fuzzy particle $C$. Fuzzification involves selecting a fuzzy function $F$ and establishing the rules for fuzzy granulation. Therefore, the process of obtaining the fuzzy particle $C$ can be expressed as follows:

$$C = F(t), t \in T \tag{2}$$

In practice, the fuzzy sets used for fuzzy information granulation typically include trapezoidal, triangular, and parabolic fuzzy sets. Herein, the membership function of the triangular fuzzy set is selected for fuzzy granulation. The functional expression is calculated as follows:

$$F(t) = \begin{cases} 0 & t < l \\ \dfrac{t-l}{z-l} & l \leq t \leq z \\ \dfrac{u-t}{u-z} & z < t \leq u \\ 0 & t > u \end{cases} \tag{3}$$

In the equation, $l$, $z$, and $u$ are the three parameters of the fuzzy information granule, corresponding to $Low$, $R$, and $Up$, respectively. Here, $Low$ represents the minimum value of the time series data variation, $R$ is the average value of the time series data variation, and $Up$ denotes the peak value of the variation in the time series data. $Low$, $R$, and $Up$ serve as the initial cluster centers (Xie et al., 2019).

**B. Fuzzy *C*-clustering**

The FCM algorithm establishes a constrained optimization model by incorporating membership degrees, enabling soft partitioning of the dataset. Herein, we primarily use the affiliation function for fuzzy *C*-means clustering. For a given dataset $T=(t_1,t_2,\cdots,t_n)$, the fuzzy rough set determines the coordinates of three initial cluster centers $C = \begin{pmatrix} c_{(1,1)} & c_{(1,2)} & c_{(1,3)} \\ \vdots & \vdots & \vdots \\ \vdots & \vdots & \vdots \\ c_{(m,1)} & c_{(m,2)} & c_{(m,3)} \end{pmatrix}$, as well as the membership degree of each sample point to all class centers $u(j,i)$, thus achieving soft partitioning of the dataset $T$:

$$u(j,i) = \dfrac{1}{\sum_p \left(d(t_i,c_{(\alpha,j)})/d(t_i,c_{(\alpha,p)})\right)^2} \quad p=1,2,3 \tag{4}$$

where $d(t_i,c_{(\alpha,j)}) = |t_i - c_{(\alpha,j)}|$ and $\mathbf{U} = [u(j,i)]$ refer to the membership matrix of the dataset $T$

and $\alpha$ is the fuzzy granule corresponding to $t_i$.

**C. Cluster center updates**

The concepts of the core and boundary regions are introduced. Let the distance from each data point $t_j$ to the cluster center $c_i$ be denoted, and let $\delta_{t_j}^{c_i}$ be the distance from the $j$-th research point to the $t$-th cluster center. To find the minimum value $\delta_j^i = min\, \delta_{t_j}^{c_i}$ from each data point to the cluster center, a threshold $\chi_1, \chi_2$ is set. $\chi_1 = r_1 \times d(t_i - \delta^i)$ and $\chi_2 = r_2 \times d(t_i - \delta^i)$, where $r_1, r_2$ are set to 0.3 and 0.7, respectively. Based on the comparison between the distance of the object to the cluster center and the minimum distance, the object points are classified into the lower approximation set, upper approximation set, and boundary region.

When the distance between object points and the cluster center is less than or equal to a specified threshold, the object point is considered to fully belong to the cluster center and is included in the upper approximation set. When the distance falls between the two thresholds, the object point is partially associated with the cluster center and is included in the lower approximation set. If the distance exceeds the second threshold, the object point is considered not to belong to the cluster center and is assigned to the boundary region:

$$\begin{aligned} &if\ \delta_j^i \leq \chi_1\ up_j^i = 1 \\ &if\ \chi_1 \leq \delta_j^i \leq \chi_2\ low_j^i = 1 \\ &if\ \chi_2 \leq \delta_j^i\ up_i = 0, low_j^i = 0 \end{aligned} \qquad (5)$$

where $up_j^i$ and $low_j^i$ represent the upper and lower approximation sets, which are represented by 0-1 matrices. An element is marked as 1 if it belongs to the corresponding set and as 0 otherwise. Additionally, the number of elements $^{num}up_j^{all} = sum(up_j^{all})$ and $^{num}low_j^{all} = sum(low_j^{all})$ in the upper and lower approximation sets corresponding to each cluster center $c_{(m,j)}$ is calculated for the given set $t_j$. The variables $\chi_1$ and $\chi_2$ denote the boundaries of the upper and lower approximation sets, respectively, while $\delta_j^i$ represents the minimum distance to the cluster center. The terms $\widehat{up}_j^i$ and $\widehat{low}_j^i$ are defined as $up_j^i \times t_j$ and $low_j^i \times t_j$, respectively. Additionally, the number of elements in the upper and lower approximation sets corresponding to each $t_j$ is calculated, serving as the basis for updating the cluster centers.

$$\begin{aligned} &if\ ^{num}up_j^{all} = 0\ c_{(m,j)} = \frac{w \times sum(\widehat{low}_j^{all})}{^{num}low_j^{all}} \\ &if\ ^{num}low_j^{all} = 0\ c_{(m,j)} = \frac{(1-w) \times sum(\widehat{up}_j^{all})}{^{num}up_j^{all}} \\ &else\ c_{(m,j)} = \frac{(1-w) \times sum(\widehat{up}_j^{all})}{^{num}up_j^{all}} + \frac{w \times sum(\widehat{low}_j^{all})}{^{num}low_j^{all}} \end{aligned} \qquad (6)$$

where $w$ denotes values within the range [0,1] and $c_{(m,j)}$ represents the features extracted from each fuzzy granule (Wang et al., 2023).

**2.3 Multiple machine learning methods**

**2.3.1 BILSTM**

The BILSTM network improves upon the LSTM by incorporating two distinct LSTM layers. This bidirectional structure allows for a more thorough understanding and representation of sequence data, thereby improving their performance in sequence modeling tasks. The structure of the BILSTM is illustrated in **Fig. 3**, along with the corresponding mathematical expression.

$$\vec{p}_t = l\left(\vec{w} \cdot x_t + \vec{v} \cdot \vec{p}_{t-1} + \vec{b}\right) \qquad (7)$$

$$\vec{p}_t = l\left(\vec{w} \cdot x_t + \vec{v} \cdot \vec{p}_{t-1} + \vec{b}\right) \tag{8}$$

$$y_t = g\left(U[\vec{p}_t * \vec{p}_t] + c\right) \tag{9}$$

where $\vec{p}$ is the output of the forward LSTM, $\vec{p}_t$ is the output of the backward LSTM, and the output state of the BILSTM is obtained by concatenating $\vec{p}$ and $\vec{p}_t$ into a matrix, which is then output. $y_t$ represents the output after the BILSTM is stacked (Shen et al., 2024).

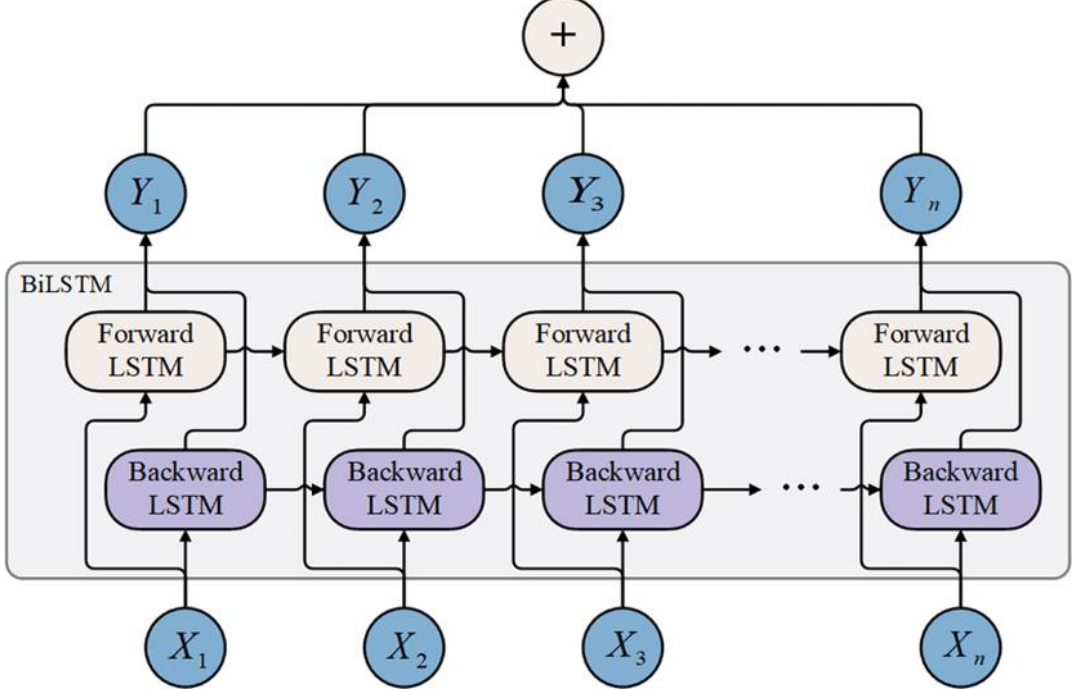

**Fig. 3 Structural diagram of the BILSTM model**

**2.3.2 CNN–GRU**

The CNN–GRU model first inputs the data through a convolution layer to extract features, then feeds them into the GRU for prediction. The GRU, a powerful variant of recurrent neural networks, is highly effective in modeling and handling sequence data (Liu et al., 2020). GRUs have demonstrated notable success in tasks such as language modeling, machine translation, audio processing, and time series analysis. The key concept of the GRU lies in its gating mechanisms, particularly those of the reset and update gates. These gates allow the GRU to manage the flow of information by selectively forgetting or retaining past hidden states while dynamically updating the hidden states based on the current input. The formulas for calculating the reset gate $q_t$ and update gate $s_t$ are as follows:

$$q_t = \sigma\left(b_q \cdot [j_{t-1}, y_t]\right) \tag{10}$$

$$s_t = \sigma\left(b_s \cdot [j_{t-1}, y_t]\right) \tag{11}$$

After the calculations, the candidate's hidden state $\tilde{j}_t$ is computed. Finally, the update gate updates the current hidden state $j_t$. The updated formula is as follows:

$$\tilde{j}_t = tanh\left(b \cdot [s_t * j_{t-1}, y_t]\right) \tag{12}$$

$$j_t = (1 - q_t) * j_{t-1} + q_t * \tilde{j}_t \tag{13}$$

### 2.3.3 LSTM–XGB

Constructing the LSTM-XGBoost prediction model involves integrating the results from several distinct individual learners using a specific strategy. This approach forms an ensemble model that offers superior prediction performance compared with that of a single individual learner. The structure of LSTM–XGB is illustrated in **Fig. 4**.

XGBoost is a tree-based boosting ensemble learning method, an enhanced version of GBDT (Kim et al., 2023). The objective function is as follows:

$$obj(\theta) = \sum_{i}^{R} l(y_i, \hat{y}_i) + \sum_{k=1}^{K} \Omega(f_k) \qquad (14)$$

The first term represents the loss function, which quantifies the overall error $y_i$ of the sample training. Here, $\hat{y}_i$ and $y_i$ denote the predicted and actual values of the load at the $i$-th point, respectively. The second term is the sum of the regularization terms for the $k$ trees.

$$\Omega(f_k) = \gamma T + \lambda \frac{1}{2} \sum_{j=1}^{T} \omega_j^2 \qquad (15)$$

where $\lambda$ is the score controlling the weight of the leaf nodes, $\omega$ is the score of the leaf nodes, and $\gamma$ is the penalty function coefficient that regulates the number of leaf nodes. The tree construction and boosting processes in the XGBoost model are guided by the objective function, with all operations focused on minimizing this objective. By incorporating a regularization term into the objective function, XGBoost effectively reduces the risk of model overfitting.

### 2.3.4 RF

The principle of random forest regression is as follows (Khan et al., 2020): assuming the training set is independently drawn from the distribution of random vectors $X$ and $Y$, let $h_i(x)$ represent the regression prediction of an individual decision tree. The final prediction of the random forest regression is then obtained by averaging the regression prediction values of all the decision trees.

$$M(X) = \frac{1}{m} \sum_{i=1}^{n} h_i(x) \qquad (16)$$

The specific process of random forest regression is as follows: for each data subset, $m$ features are randomly selected to generate a feature subset $D$:

$$D = \{j_1, j_2, \cdots, j_m\} \qquad (17)$$

Among the indices, $j_1, j_2, \cdots, j_m$ represent the indices of the $m$ randomly selected features. The $j_1, j_2, \cdots, j_m$ decision tree model is then trained using the feature subset and the corresponding data subset to obtain a predictive model, as follows:

$$f(x) = 1^T y_i * I(x \in R_i) \qquad (18)$$

Steps 1 and 2 are repeated, using $k$ decision tree models for prediction, resulting in $k$ prediction outcomes as follows:

$$f_k(x) = f(x, \theta_k) \qquad (19)$$

Among the indices, $\theta_k$ represents the parameters of the $k$-th decision tree model. The $k$ prediction results are then averaged or integrated into other ways to obtain the final regression prediction.

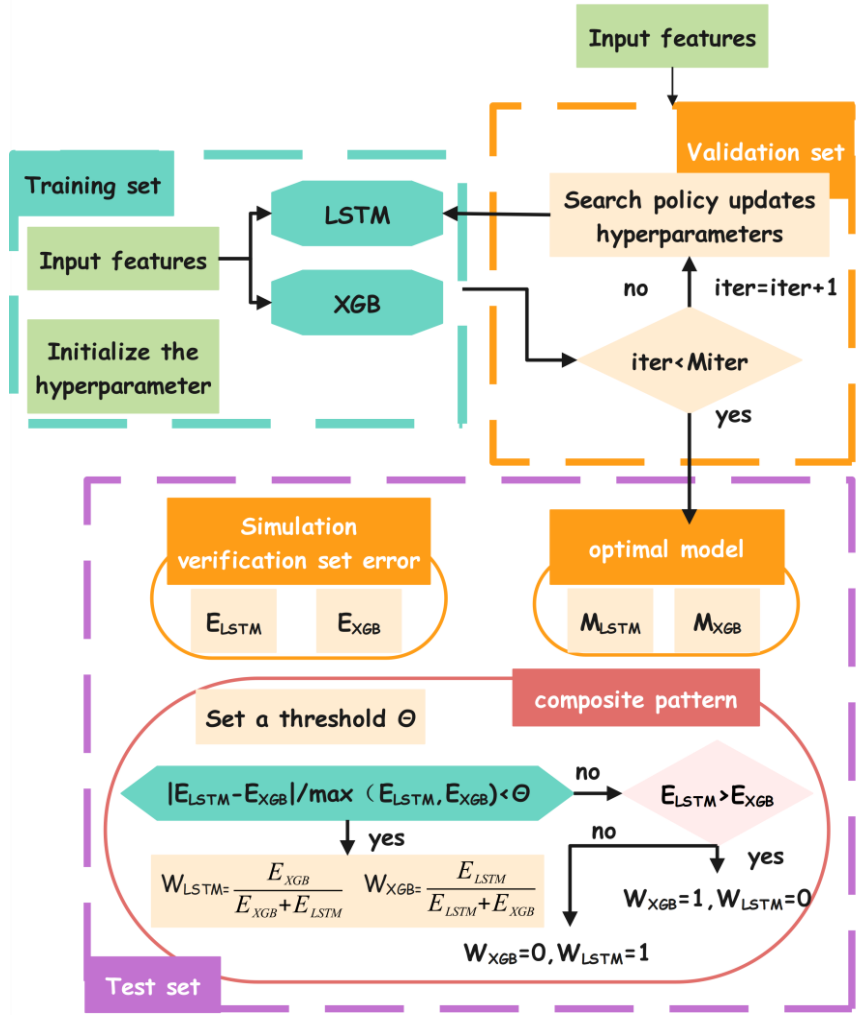

Fig. 4 Structural diagram of the LSTM–XGB model

## 2.4 IMOSFO

An alternative algorithm, IMOSFO, is proposed, in which tent mapping is incorporated during the generation of the initial population. This addition enhances the dispersion of the initial population, helping avoid local minima at the start of the computation, as shown in **Fig. 5**. Compared with random initialization, the data generated using tent mapping is more discrete, and adaptive t-distributions are applied to ensure that the sunflower moves toward the sun during the large-scale search phase at the beginning, thus preventing convergence to local minima. Later, a small-scale search is conducted to maintain search efficiency.

In this approach, an enhanced position initialization method utilizing tent mapping ensures a relatively uniform initial distribution. This technique accelerates the algorithm's convergence to an optimal solution. The uniformity of the initial distribution helps prevent the algorithm from becoming trapped in local optima, thereby improving its global search capability. The corresponding formula is as follows:

When $\Gamma(\bar{\bar{\lambda}}-1) < Apla^*$:

$$\Gamma(\bar{\bar{\lambda}}) = \frac{\Gamma(\bar{\bar{\lambda}}-1)}{Apla^*_{...}} \tag{20}$$

When $\Gamma(\bar{\bar{\lambda}}-1) >= Apla^*_{...}$:

$$\Gamma(\bar{\bar{\lambda}}) = \frac{1-\Gamma(\bar{\bar{\lambda}}-1)}{1-Apla^*_{...}} \tag{21}$$

where $\Gamma(\bar{\bar{\lambda}})$ denotes the position of the *i*-th individual and $Apla^*_{...}$ is a custom parameter.

IMOSFO is an innovative meta-heuristic algorithm that simulates the pollination process between neighboring sunflowers, guiding their movement toward the sun. As the distance between a sunflower and the sun increases, the radiation intensity weakens; conversely, as the distance decreases, the radiation intensity strengthens.

$$F = \frac{U_o}{4\pi r^2} \tag{22}$$

where $F$ represents the intensity of solar radiation, $U_o$ represents the intensity of the sun's power, and $r$ denotes the distance between the sunflower and the sun. Plants facing the sun can be represented as follows:

$$\bar{\bar{s}}_i = \frac{Y^* - Y_t}{\|Y^* - Y_t\|} t = 1,2,3,\ldots,P_n \tag{23}$$

where $Y^*$ and $Y_t$ represent the optimal and current positions of the sunflower facing the sun, respectively. $P_n$ is the number of sunflowers, and $\|\Delta\|$ denotes the norm operator. The movement of the sunflower in the direction of the sun is represented as follows:

$$d_t = \lambda \times P_t(\|Y_t + Y_{t-1}\|) \times \|Y_t + Y_{t-1}\| \tag{24}$$

where $\lambda$ represents the displacement of the sunflower and $P_t(\|Y_t + Y_{t-1}\|)$ is the pollination probability between sunflower $t$ and its adjacent sunflower $t-1$, depending on the distance between them. Populations are updated in the following manner:

$$X_{t+1} = X_t + d_t \times \bar{s}_t \tag{25}$$

where $X_{t+1}$ denotes the new location of the sunflower (Pereira & Gomes, 2023).

## 2.5 Construction of CGT-BF

In the field of wind speed prediction, integrated models are commonly used to enhance the accuracy and reliability of predictions. However, existing historical models still face significant challenges in balancing prediction stability and accuracy. To address this issue, this study proposes a bi-objective function integration model based on IMOSFO. The model effectively balances prediction stability and accuracy by simultaneously optimizing two key evaluation metrics: mean absolute percentage error (MAPE) and mean squared error (MSE).

Specifically, MAPE is used to measure the relative deviation between predicted and actual values of the model to provide a clear indication of prediction accuracy. Conversely, MSE penalizes large errors by calculating the square of the prediction error, which helps minimize large deviations in predictions. By optimizing both objectives through the IMOSFO algorithm, the integrated model achieves a strong balance between them, resulting in high prediction accuracy and good stability in practical applications.

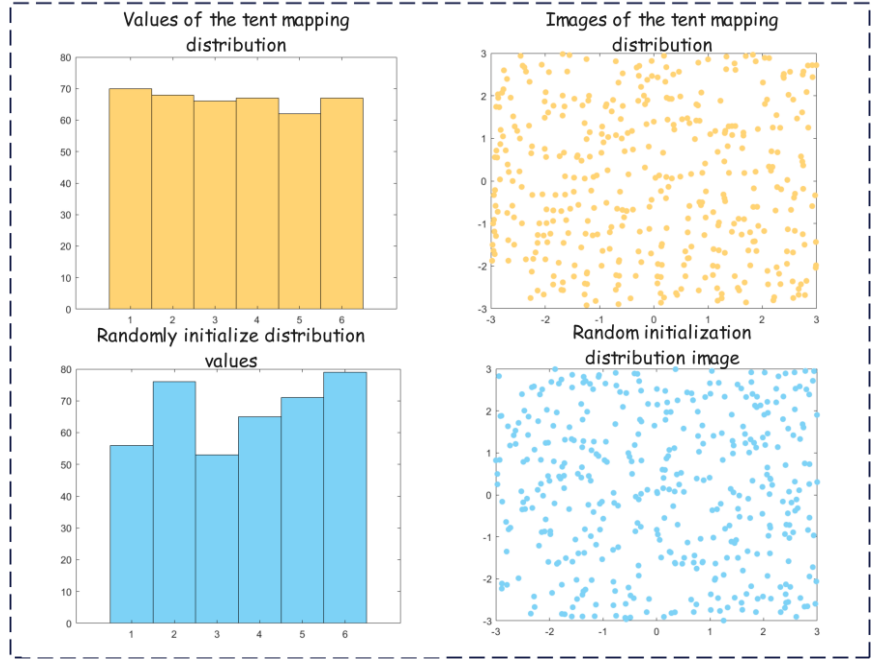

**Fig. 5 Comparison of tent mapping techniques**

## 3. Results of the four experiments

### 3.1 Experiment I

In Experiment I, we focused on the performance of the proposed method relative to those of traditional single machine learning methods for point and interval predictions and employed the proposed method for wind speed prediction. The experiment involved a simulation analysis using wind speed measurement data from Penglai, Shandong. The results indicated that the proposed model outperformed the baseline model in most cases. Specifically, we conducted a comparative analysis of CGT-BF against traditional single models, including BILSTM, CNN–GRU, LSTM–XGB, and RF. For the experiment, we used the first 60% of the data as the training set, the 60%–80% range as the validation set for parameter updates, and the final 20% as the test set for validation. The experimental results are displayed in **Table 2**.

We compared the performance of various models in both interval and point predictions. For point predictions, the parameters for $BILSTM$ were set as $B_{size}^{a}$ = 100 and $Node_{hidden}^{\sim}$ = [128, 64, 32]. The proposed model improved $Mape$ by $Mape_{Data1}^{BILSTM}$ = 0.343, demonstrating its accuracy. By contrast, the parameters for the $CNN-GRU$ model were set as $Learn_{rate}^{\sim}$ = 0.001, $Learn_{rate}^{c\sim}$ = 0.0002, and $B_{size}^{a}$ = 100, which led to a 25.2% decrease in $MSE_{Data1}^{CNN-GRU}$.

For the model $LSTM-XGB$, with parameters set to $Max_{Depth}$ = 1 and $Max_{train}$ = 750, the proposed model improved the metric $R_{Data2}^{2LSTM-XGB}$ by 0.004, highlighting its excellent fitting ability. When using random forest with parameters set at $Num_{tree}$ = 100, the proposed model performed well on the metric $IA$, with $IA_{Data3}^{RF}$ showing a 28.9% reduction in error. Analyzing all experimental data, the advantage of the proposed model in prediction accuracy was evident across the different methods and datasets. For instance, using the metric $MAPE$, the proposed model showed improvements of 0.312, 1.1490, 0.304, and 2.025 compared with those of the four original models on the metric $Data3$. Experimental results indicate that the proposed model surpasses others in all evaluated metrics, with notable improvements in prediction accuracy, model stability, and fitting capability.

For interval predictions, we used 95% and 85% confidence intervals to evaluate the generalization capability of the proposed model. For example, using the *Data2* variable, the proposed method was compared with four methods (BILSTM, CNN–GRU, LSTM–XGB, and RF) across various confidence intervals. At the $S_{Data2}^{95\%}$ confidence interval, *PICP* improved by 0.025, 0.012, 0.0, and 0.012 compared to those of BILSTM, CNN–GRU, LSTM–XGB, and RF, respectively. At the $S_{Data2}^{85\%}$ confidence interval, the proposed model showed improvements of 0.012, 0.033, and 0.004 over BILSTM, CNN–GRU, and LSTM–XGB, respectively. However, compared with the *RF* model, the proposed model on the metric *PICP* showed a slight decrease in performance, with a reduction of 0.004. In addition to *PICP*, the average interval score *AIS* is crucial. Using analysis *Data*2, at the $S_{Data2}^{95\%}$ confidence interval, the proposed model improved the average interval scores compared with BILSTM, CNN–GRU, LSTM-–XGB, and RF by 0.062, 0.107, 0.008, and 0.044, respectively; at the $S_{Data2}^{85\%}$ confidence interval, the improvements were 0.046, 0.119, 0.01, and 0.072, respectively. While the proposed model showed a slight decrease in the metric *PICP* compared with the original model, considering the more critical overall score *AIS*, it continued to demonstrate strengths in the overall performance. This suggests that the proposed model has advantages in both point forecasting and interval forecasting frameworks. The results are shown in **Fig. 6**.

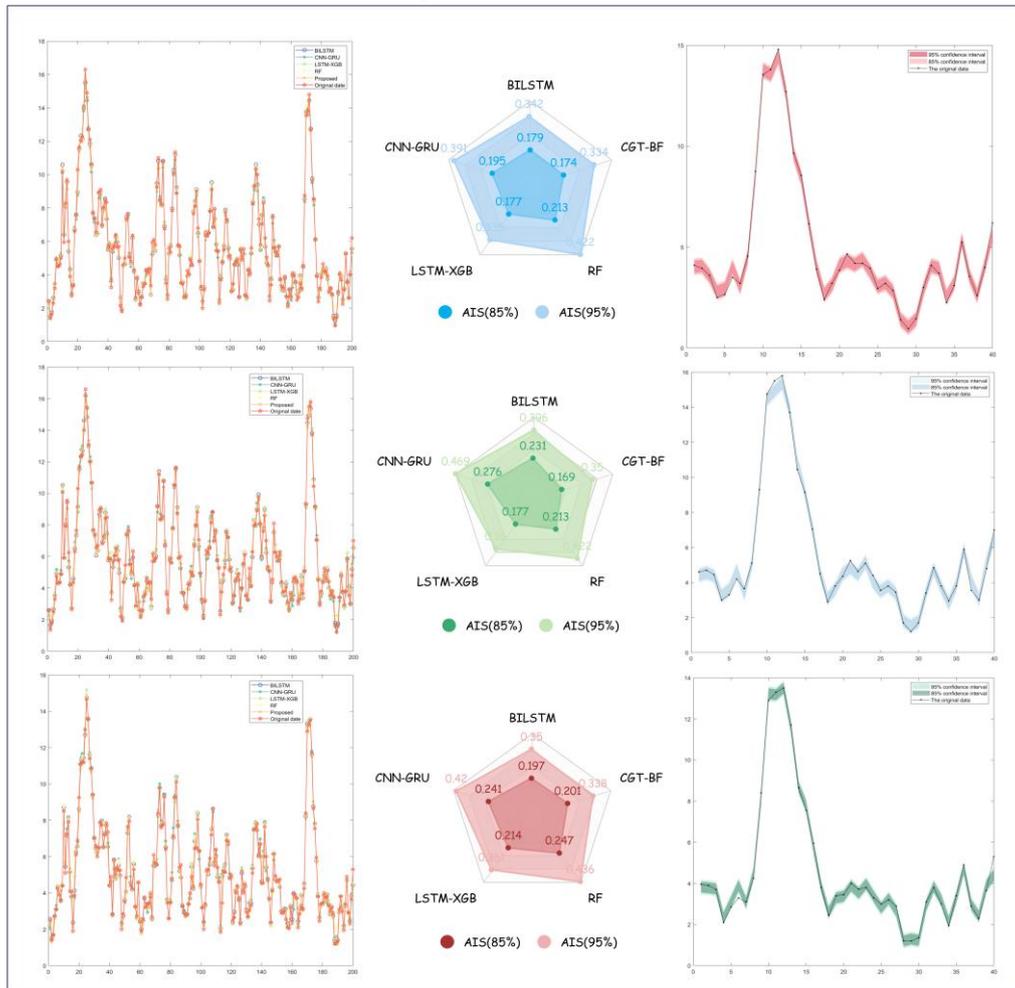

**Fig. 6 Results of Experiment I**

**Table 2**
**Results of Experiment I**

| Data | Method | MAPE | MSE | $R^2$ | PICP | | PINAW | | AIS | |
|------|--------|------|-----|-------|------|------|-------|------|-----|-----|
| | | | | | 0.95 | 0.85 | 0.95 | 0.85 | 0.95 | 0.85 |
| Data1 | GRU | 5.053 | 0.104 | 0.988 | 0.881 | 0.761 | 0.059 | 0.043 | 0.213 | 0.399 |
| | XGBOOST | 4.799 | 0.097 | 0.989 | 0.893 | 0.811 | 0.058 | 0.043 | 0.204 | 0.383 |
| | BILSTM | 4.088 | 0.079 | 0.990 | 0.897 | 0.794 | 0.054 | 0.040 | 0.179 | 0.342 |
| | CNN−GRU | 4.548 | 0.103 | 0.987 | 0.885 | 0.770 | 0.062 | 0.045 | 0.195 | 0.391 |
| | LSTM−XGB | 3.921 | 0.078 | 0.991 | 0.897 | 0.823 | 0.054 | 0.040 | 0.177 | 0.335 |
| | RF | 5.834 | 0.117 | 0.986 | 0.885 | 0.802 | 0.065 | 0.048 | 0.213 | 0.422 |
| | proposed | 3.745 | 0.077 | 0.991 | 0.901 | 0.827 | 0.054 | 0.040 | 0.174 | 0.334 |
| Data2 | GRU | 4.629 | 0.089 | 0.990 | 0.885 | 0.794 | 0.062 | 0.045 | 0.182 | 0.372 |
| | XGBOOST | 4.806 | 0.099 | 0.988 | 0.889 | 0.815 | 0.067 | 0.049 | 0.194 | 0.395 |
| | BILSTM | 3.945 | 0.078 | 0.991 | 0.872 | 0.786 | 0.054 | 0.041 | 0.231 | 0.396 |
| | CNN−GRU | 4.978 | 0.119 | 0.986 | 0.885 | 0.765 | 0.623 | 0.477 | 0.276 | 0.469 |
| | LSTM−XGB | 5.282 | 0.103 | 0.987 | 0.897 | 0.794 | 0.054 | 0.038 | 0.177 | 0.360 |
| | RF | 5.418 | 0.115 | 0.986 | 0.885 | 0.802 | 0.065 | 0.048 | 0.213 | 0.422 |
| | proposed | 3.780 | 0.077 | 0.991 | 0.897 | 0.798 | 0.058 | 0.043 | 0.169 | 0.350 |
| Data3 | GRU | 5.613 | 0.090 | 0.987 | 0.860 | 0.753 | 0.057 | 0.042 | 0.209 | 0.382 |
| | XGBOOST | 5.344 | 0.105 | 0.985 | 0.864 | 0.786 | 0.064 | 0.047 | 0.228 | 0.408 |
| | BILSTM | 4.507 | 0.080 | 0.988 | 0.881 | 0.786 | 0.054 | 0.399 | 0.197 | 0.350 |
| | CNN−GRU | 5.344 | 0.110 | 0.984 | 0.860 | 0.749 | 0.058 | 0.043 | 0.241 | 0.420 |
| | LSTM−XGB | 4.499 | 0.088 | 0.987 | 0.901 | 0.819 | 0.058 | 0.042 | 0.214 | 0.361 |
| | RF | 6.242 | 0.126 | 0.981 | 0.897 | 0.802 | 0.071 | 0.052 | 0.247 | 0.436 |
| | proposed | 4.195 | 0.077 | 0.989 | 0.905 | 0.827 | 0.053 | 0.039 | 0.201 | 0.338 |

**3.2 Experiment II**

In Experiment II, the proposed model was compared with others using different optimization algorithms, with results presented in **Table 3**. For point predictions, using $Data3$ as the metric, compared with the $MSSA$ model, the proposed model outperformed all the prediction metrics, with improvements of 0.334 in $Mape_{proposed}^{MSSA}$ and 0.012 in $MSE_{proposed}^{MSSA}$ and minor enhancements of 0.004 in $IA_{prposed}^{MSSA}$ and 0.002 in $R_{proposed}^{2MSSA}$. While the other combination models slightly underperformed compared with $proposed$, a comparison between Experiments I and II showed that most combination models considerably outperformed traditional models in point predictions. For instance, when comparing the models $MODA$ and $BILSTM$ across three datasets using the metric $Mape$, $MAPE_{MODA}^{BILSTM}$ demonstrated improvements of 0.097, 0.015, and 0.294, respectively. This clearly demonstrates that most of the combination models considerably enhanced prediction accuracy, underscoring the strengths of the proposed model. However, no single model consistently excelled across all the evaluation metrics for all the datasets. When comparing predictions with other combination models, the proposed model performed similarly in the metric $R^2$, suggesting no distinct advantage, pointing to potential areas for future improvement.

The proposed model exhibited an excelled performance in interval predictions. Specifically, for the variable $Data1$, when comparing $proposed$ to four other combination models on the metric $PICP$ within a confidence interval $S_{Data2}^{95\%}$, the $PICP$ scores were higher than those of the four models by 0.004, 0.016, 0.0, and 0.002, respectively. Within confidence interval $S_{Data2}^{85\%}$, $PICP$ outperformed the four models by 0.016, 0.012, 0.029, and 0.037, respectively. After comparing with $PICP$, the average score $AIS$ becomes a focal point for interval predictions. Analyzing the experimental data for $Data2$ within confidence interval $S_{Data2}^{95\%}$, the proposed model's scores increased by 0.02, 0.017, 0.002, and 0.012, respectively. Within confidence interval $S_{Data2}^{85\%}$, $AIS$ scores on $MODA, NSMFO, GOGWO$ improved by 0.012, 0.036, and 0.054, respectively, while scores for $MSSA$ decrease by 0.005. This indicates some variability in predictive performance across different confidence intervals. However, overall, the proposed model demonstrated excellent stability and superiority, consistently outperforming other combination models in interval prediction accuracy, particularly at high confidence levels (95%). Experimental results are shown in **Fig. 7**.

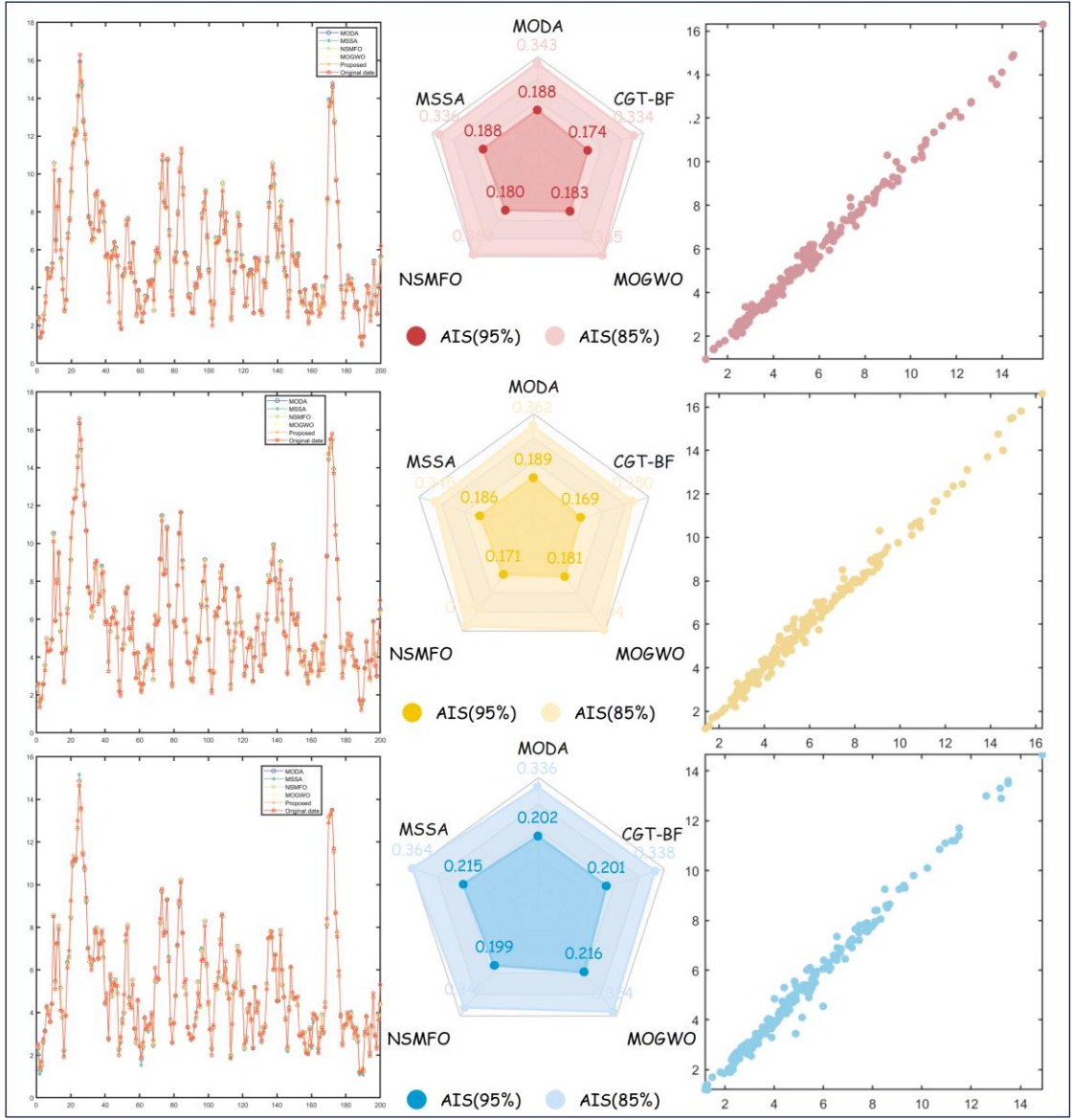

**Fig. 7 Results of Experiment II**

**Table 3**
**Results of Experiment II**

| Data | Method | MAPE | MSE | $R^2$ | PICP | | PINAW | | AIS | |
|---|---|---|---|---|---|---|---|---|---|---|
| | | | | | 0.95 | 0.85 | 0.95 | 0.85 | 0.95 | 0.85 |
| Data1 | MODA | 3.991 | 0.078 | 0.991 | 0.897 | 0.811 | 0.053 | 0.039 | 0.188 | 0.343 |
| | MSSA | 4.083 | 0.079 | 0.991 | 0.885 | 0.815 | 0.053 | 0.041 | 0.188 | 0.336 |
| | NSMFO | 4.158 | 0.084 | 0.990 | 0.901 | 0.798 | 0.057 | 0.042 | 0.180 | 0.360 |
| | MOGWO | 3.862 | 0.081 | 0.990 | 0.889 | 0.790 | 0.055 | 0.039 | 0.183 | 0.365 |
| | Proposed | 3.745 | 0.077 | 0.991 | 0.901 | 0.827 | 0.054 | 0.040 | 0.174 | 0.334 |
| Data2 | MODA | 3.930 | 0.077 | 0.991 | 0.893 | 0.782 | 0.058 | 0.043 | 0.189 | 0.362 |
| | MSSA | 4.175 | 0.079 | 0.991 | 0.893 | 0.811 | 0.059 | 0.045 | 0.186 | 0.345 |
| | NSMFO | 3.946 | 0.082 | 0.990 | 0.893 | 0.798 | 0.062 | 0.043 | 0.171 | 0.386 |
| | MOGWO | 4.537 | 0.093 | 0.989 | 0.885 | 0.798 | 0.062 | 0.043 | 0.181 | 0.404 |
| | Proposed | 3.780 | 0.077 | 0.991 | 0.897 | 0.798 | 0.058 | 0.043 | 0.169 | 0.350 |
| Data3 | MODA | 4.213 | 0.078 | 0.989 | 0.914 | 0.827 | 0.053 | 0.039 | 0.202 | 0.336 |
| | MSSA | 4.529 | 0.089 | 0.987 | 0.889 | 0.802 | 0.056 | 0.041 | 0.215 | 0.364 |
| | NSMFO | 4.256 | 0.077 | 0.989 | 0.897 | 0.811 | 0.053 | 0.039 | 0.199 | 0.340 |
| | MOGWO | 4.250 | 0.077 | 0.989 | 0.881 | 0.807 | 0.051 | 0.038 | 0.216 | 0.354 |
| | Proposed | 4.195 | 0.077 | 0.989 | 0.905 | 0.827 | 0.053 | 0.039 | 0.201 | 0.338 |

### 3.3 Experiment III

The objective of Experiment III was to validate the generalization ability of the composite model proposed in this study for forecasting. To achieve this, we used a five-fold cross-validation method to redistribute and evaluate the data. The results are detailed in **Table 4**. In the five-fold cross validation, the proposed composite model exhibited high prediction accuracy and stability. For instance, with Data1, the MAPE values were $MAPE_{Data1}^{flod-1} = 4.178$, $MAPE_{Data1}^{flod-2} = 4.327$, $MAPE_{Data1}^{flod-3} = 4.082$, $MAPE_{Data1}^{flod-4} = 4.204$, and $MAPE_{Data1}^{flod-5} = 4.693$, maintaining an average of approximately 4.1, suggesting stable predictive performance across different training and test set distributions. For Data2, the coefficient of determination ($R^2$) was $R_{Data2}^{2,flod-1} = 0.985$, $R_{Data2}^{2,flod-2} = 0.987$, $R_{Data2}^{2,flod-3} = 0.986$, $R_{Data2}^{2,flod-4} = 0.987$, and $R_{Data2}^{2,flod-5} = 0.981$, maintaining an average of approximately 0.985, indicating stable explanatory power. With Data3, the MSE values were $MSE_{Data3}^{flod-1} = 0.112$, $MSE_{Data3}^{flod-2} = 0.131$, $MSE_{Data3}^{flod-3} = 0.100$, $MSE_{Data3}^{flod-4} = 0.101$, and $MSE_{Data3}^{flod-5} = 0.114$, maintaining an average of approximately 0.11, demonstrating consistent error control across distributions. These results highlight the composite model's strong stability and predictive performance across different datasets and metrics. They further validate the model's generalization capability and reliability, providing robust evidence of its efficacy in practical applications. However, it is important to note that while cross validation is an effective method for testing generalization ability, its high computational demand should not be overlooked. The specific results of the experiment are shown in **Fig. 8**.

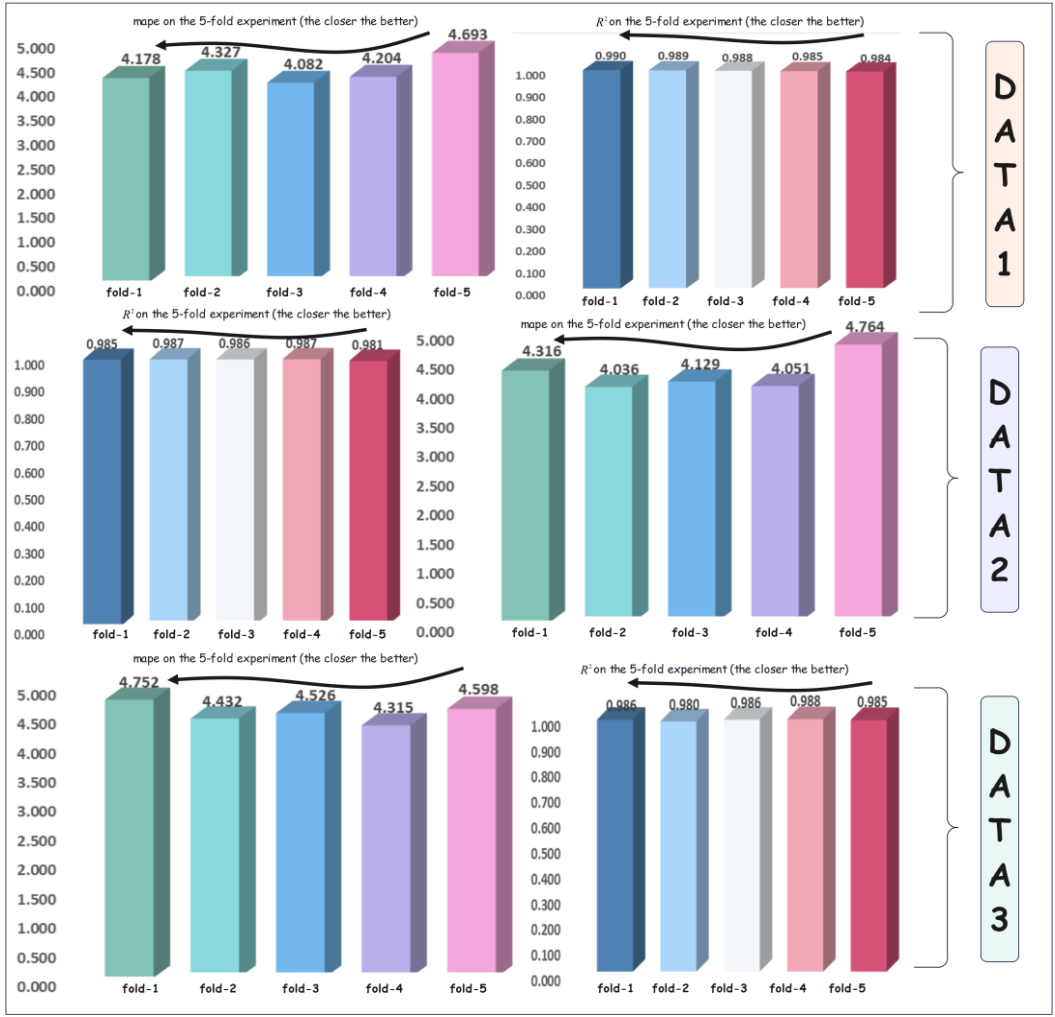

**Fig. 8 Results of Experiment III**

**3.4 Experiment IV**

In Experiment IV, we tested the optimization system using classic test functions to validate the improvement capabilities of its optimization algorithms. Specifically, we evaluated enhanced MOSFO, MSSA, and MODA using the classic test functions ZDT1, ZDT2, and ZDT3. The MOSFO parameters were $Num_{iter}^{MOSFO} = 100$ and $Dim = 4$, and the MSSA parameters were $Max_{size}^{MODA} = 100$. The ZDT functions are detailed in **Table 5**, while the specific performances of the Pareto frontier are shown in **Fig. 9**. These test functions are widely used benchmarks in the field of optimization and are effective for evaluating the performance of optimization algorithms in handling various objective functions and constraints. By testing these functions, we gained a comprehensive understanding of the optimization system's performance in different scenarios, thereby validating the ability of the improved algorithms to enhance optimization efficiency and accuracy. The experimental results provide a strong foundation for further the development and application of optimization algorithms.

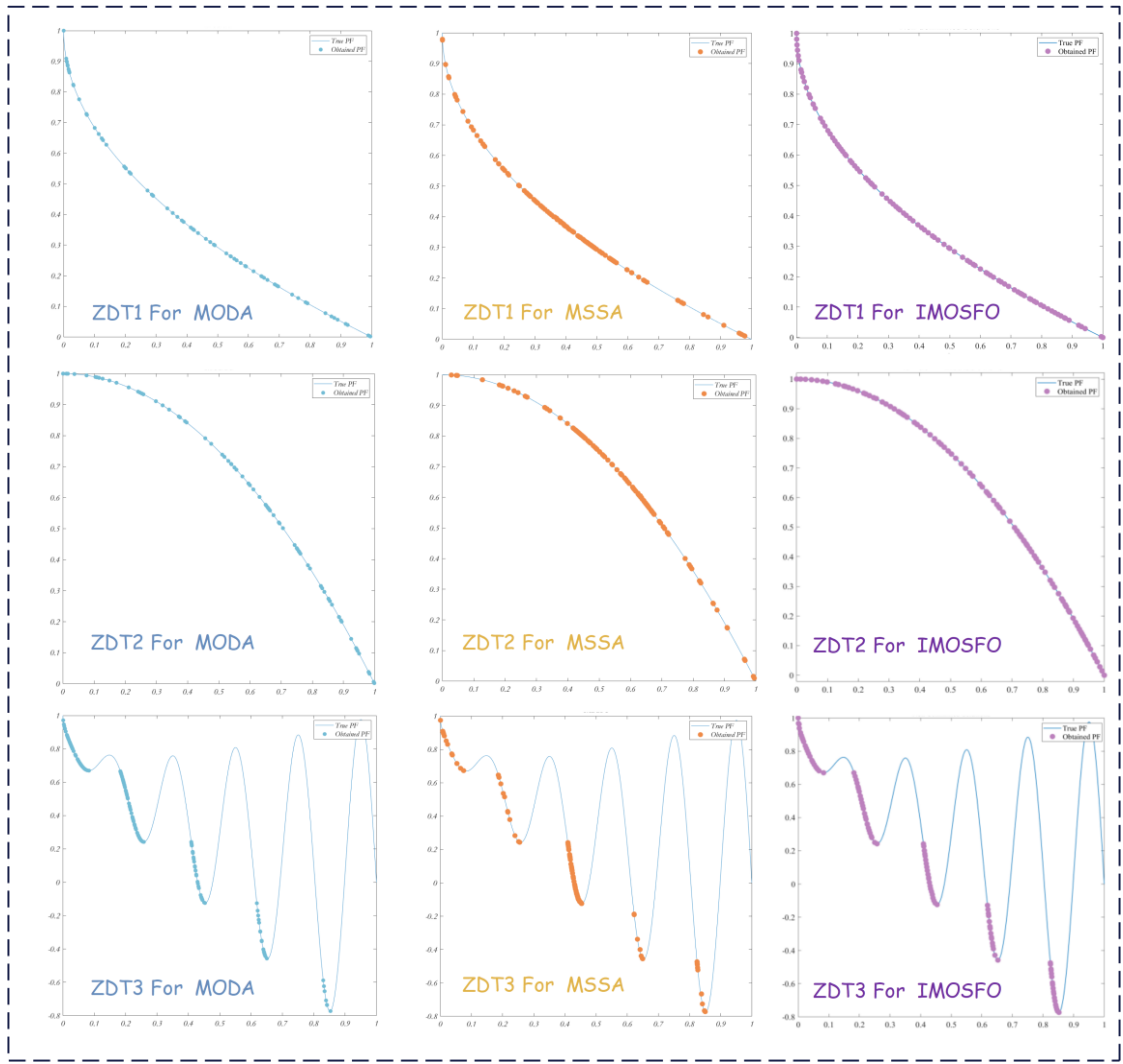

**Fig. 9 Comparison of various multiobjective optimization algorithms**

**Table 4**
**Results of Experiment III**

|  |  | MAPE | MSE | MAE | RMSE | NMSE | U1 | IA | $R^2$ |
|---|---|---|---|---|---|---|---|---|---|
| $Data1$ | $Flod-one$ | 4.178 | 0.102 | 0.221 | 0.319 | 0.010 | 0.043 | 0.958 | 0.990 |
|  | $Flod-two$ | 4.327 | 0.107 | 0.227 | 0.327 | 0.010 | 0.045 | 0.957 | 0.989 |
|  | $Flod-three$ | 4.082 | 0.103 | 0.220 | 0.321 | 0.012 | 0.047 | 0.953 | 0.988 |
|  | $Flod-four$ | 4.204 | 0.117 | 0.243 | 0.342 | 0.014 | 0.049 | 0.945 | 0.985 |
|  | $Flod-five$ | 4.693 | 0.128 | 0.253 | 0.358 | 0.015 | 0.052 | 0.944 | 0.984 |
| $Data2$ | $Flod-one$ | 4.316 | 0.126 | 0.258 | 0.355 | 0.014 | 0.049 | 0.944 | 0.985 |
|  | $Flod-two$ | 4.036 | 0.119 | 0.234 | 0.344 | 0.013 | 0.050 | 0.951 | 0.987 |
|  | $Flod-three$ | 4.129 | 0.126 | 0.240 | 0.355 | 0.013 | 0.048 | 0.951 | 0.986 |
|  | $Flod-four$ | 4.051 | 0.094 | 0.221 | 0.306 | 0.013 | 0.045 | 0.951 | 0.987 |
|  | $Flod-five$ | 4.764 | 0.151 | 0.260 | 0.389 | 0.019 | 0.057 | 0.944 | 0.981 |
| $Data3$ | $Flod-one$ | 4.752 | 0.112 | 0.239 | 0.334 | 0.013 | 0.051 | 0.949 | 0.986 |
|  | $Flod-two$ | 4.432 | 0.131 | 0.233 | 0.362 | 0.019 | 0.057 | 0.942 | 0.980 |
|  | $Flod-three$ | 4.526 | 0.100 | 0.228 | 0.316 | 0.014 | 0.049 | 0.946 | 0.986 |
|  | $Flod-four$ | 4.315 | 0.101 | 0.220 | 0.317 | 0.012 | 0.047 | 0.952 | 0.988 |
|  | $Flod-five$ | 4.598 | 0.114 | 0.229 | 0.338 | 0.014 | 0.053 | 0.948 | 0.985 |

**Table 5**
**Three benchmark test functions and their formulas**

| Test function | Formula |
|---|---|
| **ZDT1** | $G(v) = 1 + (9/(m-1))\sum v_i$ <br> $H(p_1, G) = 1 - \sqrt{p_1/G}$ |
| **ZDT2** | $G(v) = 1 + (9/(m-1))\sum v_i$ <br> $H(p_1, G) = 1 - (p_1/G)^2$ |
| **ZDT3** | $G(v) = 1 + (9/(m-1))\sum v_i$ <br> $H(p_1, G) = 1 - \sqrt{p_1/G} - (p_1/G)\sin(10\pi p_1)$ |

$$\min p_1(v) = v_1$$

$$\min p_2(v) = G(v) \cdot H(p_1(v), G(v))$$

## 4. Discussion

Through validation in the four experiments, we clearly demonstrated the substantial advantages of the proposed method in both point and interval predictions. Additionally, we highlighted the broad applicability of the model across various scenarios and validated the superiority of the improved optimization algorithm. This section is structured as follows: Section 4.1 provides a comprehensive discussion of the parameter settings used in the experiments, along with a detailed analysis of their effects on the outcomes. Section 4.2 validates the point prediction MAPE and MSE using the Diebold–Mariano (DM) test, comparing the experimental results from CGT-BF with those of other models. Section 4.3 employs the improvement rate index (IRI) for a quantitative analysis of the experimental data, further emphasizing the advantages of the proposed method.

### 4.1 Experimental parameter settings

Herein, we employed a method known as **FIC-MG** for the four experiments, with a time window of 36 in the experimental design. The proposed model was based on the sunflower multiobjective optimization algorithm, an advanced optimization technique. For the parameter settings, we set the dimensionality (*dim*) of the algorithm to 4 and used a four-dimensional weight vector ($w_1, w_2, w_3, w_4$) to drive the optimization process. Additionally, we set the maximum iterations for ISOMFO to 100 to ensure sufficient iteration opportunities for deep optimization. A critical parameter in this study was the archive size of the algorithm, which was set to 100. During the execution of the optimization algorithm, the archive size plays a crucial role in determining the number of search agents (i.e., potential solutions) available. Appropriately configuring the number of search agents is vital for both the efficiency and the quality of the optimization process. If the number of search agents is too high, this can increase the diversity of solutions; however, it will also considerably raise the computational load of the algorithm, reducing the overall efficiency. By contrast, a too-low archive size reduces computation time but may result in an insufficient search process, potentially compromising the quality of the optimization results. To verify this, we conducted a series of experimental analyses using multiple datasets with varying characteristics. The experimental results indicate that with an archive size of 100, the algorithm can achieve the optimal results while maintaining operational efficiency. This finding highlights the importance of accurately configuring the number of search agents in algorithm design. Detailed parameter settings and experimental results are provided in **Table 6**, which serves as the foundation for further analysis and discussion.

**Table 6**
**Parameters used in the four experiments**

| Model | Parameters | Setting Value |
|---|---|---|
| ***BILSTM*** | Learning rate | 0.001 |
| | batchsize | 100 |
| | Number of neurons | [128,64,32] |
| ***CNN–GRU*** | Number of neurons | [128,64,32] |
| | Learning rate | 0.001 |
| | batchsize | 150 |
| ***LSTM–XGB*** | Maximum training | 750 |
| | Learning rate | 0.001 |
| ***RF*** | Number of trees | 100 |
| ***FIC-MG*** | Upper threshold | 0.7 |
| | Lower threshold | 0.3 |
| | Center of clustering | 3 |
| ***MOGWO&MSSA&MODA&NSMFO*** | Number of objective functions | 2 |
| | Archive max size | 100 |
| | dim | 4 |
| | ub | 2 |
| | lb | $-2$ |
| ***IMOSFO*** | Number of objective functions | 2 |
| | Number of grids in each dimension | 30 |
| | Maximum number of solutions in PF | 100 |
| | Archive max size | 100 |
| | dim | 4 |
| | ub | 2 |
| | lb | $-2$ |
| | Pollination rate | 0.1 |
| | Mortality rate | 0.1 |

**4.2 DM test**

The core objective of the DM test is to analyze and compare forecasting errors from different models to determine whether any differences in performance are statistically significant. This test is particularly valuable for assessing the robustness of the experimental results, providing a reliable measure of model accuracy. Herein, the CGT-BF point forecasting system serves as the primary model, and its performance was compared against a reference model.

To begin, let us define the prediction outputs of model $\alpha$, denoted as $F_{(\alpha \to DM)}^{\zeta} = [P_{(\alpha \to DM)}^{1}, P_{(\alpha \to DM)}^{2}, \cdots, P_{(\alpha \to DM)}^{\zeta}]$, and those of the baseline model $\beta$, denoted as $K_{(\beta \to DM)}^{\zeta} = [Y_{(\beta \to DM)}^{1}, Y_{(\beta \to DM)}^{2}, \cdots, Y_{(\beta \to DM)}^{\zeta}]$. The actual observed values are represented by

$B^{\zeta}_{(true)} = [T^1_{(true)}, T^2_{(true)}, \cdots, T^{\zeta}_{(true)}]$.

For model $\alpha$, the error is $E^{(\zeta)}_{(\alpha)} = \left| F^{\zeta}_{(\alpha \to DM)} - B^{\zeta}_{(true)} \right|$, and for model $\beta$, it is $E^{(\zeta)}_{(\beta)} = \left| K^{\zeta}_{(\beta \to DM)} - B^{\zeta}_{(true)} \right|$. The original and alternative assumptions are as follows:

$$H_0 : \sum_{\zeta=1}^{\omega} \left\{ \left( E^{(\zeta)}_{(\alpha)} \right)^2 - \left( E^{(\zeta)}_{(\beta)} \right)^2 \right\} = 0 \tag{26}$$

$$H_1 : \sum_{\zeta=1}^{\omega} \left\{ \left( E^{(\zeta)}_{(\alpha)} \right)^2 - \left( E^{(\zeta)}_{(\beta)} \right)^2 \right\} \neq 0 \tag{27}$$

The test statistic is computed using the following formula:

$$\pi^{\oplus}_{DM} = \frac{\sum_{\zeta=1}^{\omega} \left\{ \left( E^{(\zeta)}_{(\alpha)} \right)^2 - \left( E^{(\zeta)}_{(\beta)} \right)^2 \right\}}{\sqrt{\omega} * \sum_{\zeta=1}^{\omega} \left\{ E^{(\zeta)}_{(\alpha)} - E^{(\zeta)}_{(\beta)} \right\}} \tag{28}$$

At a significance level of $\alpha = 0.05$, the rejection criterion is defined as $W = \left| \pi^{\oplus}_{DM} \right| > \left| z_{\alpha/2} \right|$. If the calculated statistic lies within the rejection region, the null hypothesis is rejected, indicating a statistically significant difference in the forecasting performance between the two models. The experimental results are summarized in **Table 7**.

**A.** For Data3, the DM test value for the **BILSTM** on MAPE is $DM^{Bilstm}_{Site3} \to Mape = 13.51$, suggesting that the prediction performance of the **CGT-BF** model is significantly better than that of **BILSTM**. The experimental results also demonstrate that **CGT-BF** has a significant advantage over the **RF** model, with DM test results on MAPE of $DM^{RF}_{Site1} \to Mape = 12.101$, $DM^{RF}_{Site2} \to Mape = 12.077$, and $DM^{RF}_{Site3} \to Mape = 13.225$, and DM test results on MSE of $DM^{RF}_{Site1} \to MSE = 9.146$, $DM^{RF}_{Site2} \to MSE = 8.426$, and $DM^{RF}_{Site3} \to MSE = 8.46$. The results highlight that the proposed model outperforms the other models in all metrics, particularly in prediction accuracy, model stability, and fitting capability.

**B.** The DM test values for the BILSTM, CNN–GRU, LSTM–XGB, RF, MOGWO, MSSA, MODA, and NSMFO models on MAPE were all positive, indicating that the proposed CGT-BF prediction system achieved higher prediction accuracy than the four benchmark models and the four combined models.

**Table 7**
**Results of the DM test**

|  | Data1 | | Data2 | | Data3 | |
| --- | --- | --- | --- | --- | --- | --- |
|  | MAPE | MSE | MAPE | MSE | MAPE | MSE |
| *BILSTM* | 12.017 | 9.173 | 11.866 | 8.426 | 13.510 | 8.386 |
| *CNN – GRU* | 12.209 | 9.200 | 12.129 | 8.466 | 13.419 | 8.420 |
| *LSTM – XGB* | 12.043 | 9.183 | 11.983 | 8.430 | 13.586 | 8.396 |
| *RF* | 12.101 | 9.146 | 12.077 | 8.426 | 13.225 | 8.460 |
| *MODA* | 2.415 | −0.336 | 2.147 | −0.784 | 0.552 | 0.410 |
| *MSSA* | 3.416 | 0.660 | 1.957 | −0.838 | 1.777 | 2.988 |
| *NSMFO* | 3.096 | 1.124 | 1.919 | 1.672 | 0.627 | −0.068 |
| *MOGWO* | 1.606 | 1.167 | 3.958 | 2.689 | 0.620 | −0.265 |

**4.3 Index improvement rate**

MAPE is widely recognized as a key indicator of prediction accuracy in point predictions. To further evaluate the accuracy of point predictions, we introduced the IRI as a core assessment tool. Specifically, we calculated MAPE to measure the enhancement in point prediction results. Building on this, we developed an improvement rate formula that quantitatively assessed the performance boost of the prediction model under various conditions. This formula provides a valid criterion to evaluate the effectiveness of the model's predictions after optimization.

$$IRI_{Mape} = \left| \frac{Mape_{CGT-BF} - Mape_{O}}{Mape_{CGT-BF}} \right| \times 100\% \tag{29}$$

Herein, we tested the CGT-BF model, and the final validation results are presented in **Table 8**. A detailed analysis based on these results is as follows.

**A.** First, we compared the performance of the CGT-BF model with the four basic models (BILSTM, CNN–GRU, LSTM–XGB, and RF) in terms of the IRI. The findings revealed that the CGT-BF model considerably outperformed the basic models across all three datasets. Specifically, compared with the BILSTM model, the CGT-BF model achieved an average improvement rate of $IRI_{CGT-BF}^{BILSTM} \rightarrow (MAPE) =$ 6.99%. Against the CNN–GRU model, the IRI was $IRI_{CGT-BF}^{CNN-GRU} \rightarrow (MAPE) = 26.836\%$, while it was $IRI_{CGT-BF}^{LSTM-XGB} \rightarrow (MAPE) = 17.232\%$ against the LSTM–XGB model, and a remarkable $IRI_{CGT-BF}^{RF} \rightarrow (MAPE) = 49.305\%$ against the RF model. These results demonstrate the substantial advantages of the CGT-BF model in handling relevant tasks over traditional models, highlighting its effectiveness in enhancing predictive performance and addressing complex data structures.

**B.** We further compared the CGT-BF model with four composite models (MODA, MSSA, NSMFO, and MOGWO) in terms of the IRI. The CGT-BF model showed an average improvement rate of $IRI_{CGT-BF}^{MODA} \rightarrow (MAPE) = 3.650\%$ over the MODA model across all datasets, with particularly better performance on datasets 1 and 2, although its improvement was more limited on dataset 3. Compared with the MSSA model, the CGT-BF model showed a significant average improvement of $IRI_{CGT-BF}^{MSSA} \rightarrow (MAPE) = 9.146\%$, outperforming the MSSA model on all datasets, especially on dataset 2. When compared with the NSMFO model, the CGT-BF model achieved an average improvement of $IRI_{CGT-BF}^{NSMFO} \rightarrow (MAPE) = 5.625\%$, with the most notable improvement on dataset 1. Compared with the MOGWO model, the CGT-BF model exhibited an improvement rate of $IRI_{CGT-BF}^{MOGWO} \rightarrow (MAPE) = 8.160\%$, particularly on dataset 2, where the improvement was most pronounced, while the improvement was relatively less on datasets 1 and 3.

In summary, the CGT-BF model consistently demonstrated superior performance compared with both traditional and composite models (BILSTM, CNN–GRU, LSTM–XGB, RF, MODA, MSSA, NSMFO, and MOGWO). This improvement was not limited to any specific dataset but was observed across multiple datasets, underscoring the adaptability and superiority of the CGT-BF model for handling diverse data types.

**Table 8**
**Improvement rate of point forecast indices**

|  | BILSTM | CNN–GRU | LSTM–XGB | RF | MODA | MSSA | NSMFO | MOGWO |
|---|---|---|---|---|---|---|---|---|
| Data1 | 9.179 | 21.456 | 4.715 | 55.794 | 6.567 | 9.046 | 11.038 | 3.147 |
| Data2 | 4.366 | 31.687 | 39.736 | 43.333 | 3.957 | 10.446 | 4.403 | 20.027 |
| Data3 | 7.424 | 27.366 | 7.246 | 48.789 | 0.427 | 7.945 | 1.435 | 1.305 |

## 5. Conclusion

Wind energy is inherently variable, making accurate wind speed forecasting crucial for wind farm operation and grid safety. However, many existing studies often overlook effective data preprocessing methods and adaptive forecasting strategies, leading to suboptimal forecast accuracy. To address these challenges, this study introduces the CGT-BF system for wind speed forecasting, which integrates advanced data extraction techniques with neural networks. The system uses an FIC-MG preprocessing method based on fuzzy set theory to extract eigenvalues from wind speed data. Additionally, an improved sunflower optimization algorithm is employed, and the neural network is optimized using a multiobjective meta-heuristic algorithm. Results from Experiments I and II demonstrate that the proposed CGT-BF prediction system substantially enhances wind speed prediction accuracy, achieving an average point prediction improvement of 21.6%. Experiment III confirms the robust generalization capability of the proposed method, with consistent results observed across a five-fold cross validation. Experiment IV highlights that the proposed IMOSFO multiobjective optimization algorithm outperforms other optimization algorithms in terms of search efficiency.

The self-adaptive prediction system presented in this study not only enables more accurate wind speed prediction but also expands the scope of wind speed prediction research. However, there are areas for further improvement. First, conducting a multivariate analysis that includes additional meteorological conditions, such as air temperature and barometric pressure, could enhance prediction accuracy. Second, strengthening the ability to analyze long-term wind speed data is crucial for better understanding of wind speed variation patterns and improving forecasting accuracy.

While the dual-frame prediction system proposed in this study offers high accuracy, it slightly increases the model's complexity. Future research could explore the balance between model complexity and prediction accuracy. Additionally, short-term multistep wind speed prediction could be considered as a direction for future research to validate the effectiveness of the proposed method.